\documentclass{elsart5p}
\usepackage[english]{babel}
\usepackage{graphicx}
\usepackage{latexsym}
\usepackage{amssymb}
\usepackage{amsmath}

\journal{Elsevier NIM A}

\begin{document}

\begin{frontmatter}

\title{Micro Drift Chamber as a precise vertex detector 
for the DIRAC experiment.}

\author{A.V. Dudarev\corauthref{cor1}\thanksref{label1}},
\corauth[cor1]{Corresponding author.}
\ead{Andrey.Dudarev@jinr.ru}
\thanks[label1]{Supported by INTAS YSF Ref. Nr. 05-109-5080}
\author{V.V. Kruglov},
\author{L.Yu. Kruglova},
\author{M.V.Nikitin}

\address{Joint Institute for Nuclear Research, Dubna, Russia}

\begin{abstract}

A possible implementation of the Micro Drift Chamber (MDC) technique 
as a high resolution vertex detector in the upstream part of the DIRAC 
spectrometer was investigated in this paper. Simulations of different 
MDC layout were performed with the help of GARFIELD package. Based on 
the results of simulation the optimal chamber geometry and the gas 
mixture were selected. 

One cell prototype was produced and its characteristics were measured 
at different particle fluxes, various gas pressures and thresholds of 
electronics. 

Data observed together with expected coordinate resolution of MDC allow 
to employ such detector in different field of application including the 
DIRAC experiment.
\end{abstract}

\date{}

\end{frontmatter}

\section{Introduction}
\begin{figure*}[ht]
\begin{center}
  \includegraphics[width=14.5cm]{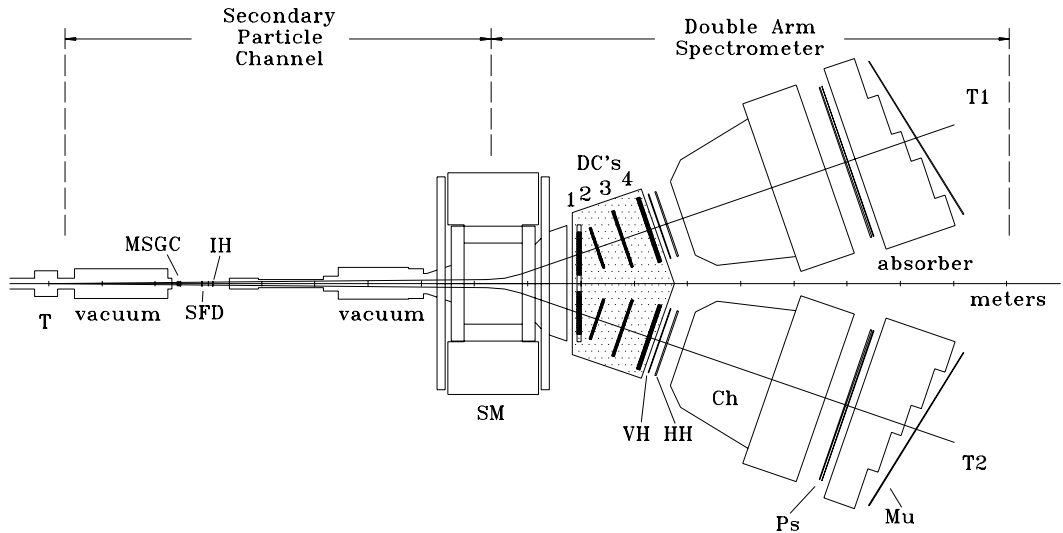} 
  \caption{Experimental setup of the DIRAC experiment: $T$ - target
station, $MSGC$ - micro-strip gas chambers, $SFD$ - scintillating 
fiber detector, $IH$ - ionization hodoscopes, $SM$ - spectrometer magnet, 
$DC$ - drift chamber system, $VH$ - vertical hodoscopes, $HH$ - horizontal 
hodoscopes, $Ch$ - Cherenkov counters, $Ps$ - preshower detectors, $Mu$ - muon
scintillation counters, $T1$ and $T2$- spectrometer arms.} 
  \label{fig:dirac}
\end{center}
\end{figure*}

The DIRAC experiment~\cite{DIRAC first_res} aims to measure with high 
precision the lifetime of $\pi^{+}\pi^{-}$ atoms as well as atoms consisted 
of charged $\pi$ and $K$-mesons~\cite{DIRAC addendum}. The experiment is 
carrying out at CERN Proton Synchrotron on the extracted $24~\text{GeV}/c$ 
proton beam. The DIRAC setup (fig.~~\ref{fig:dirac})~\cite{DIRAC Setup} is 
a magnetic double arm spectrometer. It consists of a proton beam line, 
target station, secondary particles channel, spectrometer magnet and 
upstream and downstream detectors including 28 planes of the Drift Chambers 
as a main tracking system.

Events investigated in the experiment are characterized by pairs of particles
with small relative momentum ($Q<3~\text{MeV}/c$). In the upstream part of the 
spectrometer the distance between such particles is $\sim200~\mu\text{m}$.
For their registration the MSGC, SFD and IH detectors are used. But the
absence of a rigorous tracking system, which is able to register reliably
the close tracks, affects the experiment efficiency. Therefore it seems 
quite reasonable to supplement the experimental setup with such system.
Amount of information from these detectors is also extremely large and 
makes difficulties for data acquisition and analysis. Traditional solid-state
and gaseous multi-channel vertex detectors are universal instruments
appropriate for experiments with high particle multiplicity. At once there
are many experiments where multiplicity is not so high, but it is necessary
to detect pairs of the close particles and measure their coordinates with 
high accuracy. The DIRAC setup belongs to this type of experiment.

Aim of the present work is investigation of the possibility to develop a
vertex detector based on conventional technique of the drift chamber with
small drift distance - micro drift chamber (MDC).

Next in the paper we describe the principle of this detector operation,
which thoroughly enables to meet the requirements of the DIRAC and similar 
experiments. The results of the detector properties simulation using 
Garfield package are presented. Some data obtained in experimental 
investigation of one-cell prototype of the MDC are shown.

\section{Features of two close tracks registration in drift chamber}
Let us consider how particles are detected in a drift chamber cell.
In case of one particle primary ionization electrons drift to the anode 
wire and initiate the avalanche process. This avalanche occupies part 
of the anode wire and inhibits for some time the detection of another 
particle close-by in space. Therefore, if two particles cross the cell
simultaneously, one of them could not be detected because either it does 
not produce an avalanche or due to electronics dead time. 

\begin{figure}[ht]
\begin{center}
  \includegraphics[width=8.5cm]{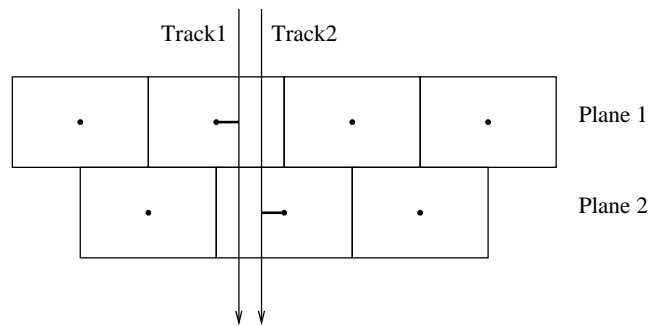} 
  \caption{Scheme of close tracks registration by two shifted planes of
the drift chamber.} 
  \label{fig:2planes}
\end{center}
\end{figure}

These restrictions are important in case of using one plane of the
drift chamber. Complementary plane shifted by half a cell width, which
is usually employed to eliminate a left-right ambiguity, in 
addition allows to resolve the problem of two close tracks 
registration (fig.~\ref{fig:2planes}). As one can see in the figure,
if two tracks cross one cell, then the signal in each plane will
be produced by that particle, which is closer to the wire. Therefore
both particles will be registered.

Using of double planes chamber besides the possibility of two close
tracks registration enables also interesting opportunity to organize
fast and effective trigger for such events selection. If one particle
cross two adjacent cells the sum of the drift times for these cells 
is equal to the maximum drift time. In case of two tracks this sum 
will decrease with increasing the distance between these tracks.
If we put the signals from these cells to the unit, which operates 
as a mean-timer, then we will get a time spectrum with the peak
corresponding to the case of one particle and the rest part of the
spectrum corresponds to more than one particle. Then if we gate the 
signals from mean-timers through the time window one can to select 
the required events on the fly. Efficiency of this trigger and its 
possible realization in the DIRAC experiment were represented 
in \cite{trigger}.

\section{Choice of MDC geometric parameters and operation mode}
Lateral dimension of the secondary particle beam in the supposed MDC region 
is $\sim80\times80~\text{mm}^{2}$ and particle flux is
$\sim10^6~\text{sec}^{-1}~\text{cm}^{-2}$. It is necessary to minimize the
probability of tracks overlaying from successive interactions of beam 
particle with target. Taking into account that number of channels in one 
MDC plane should be multiple of 16 and one cell dimension should be small 
enough to distinguish efficiently two close tracks, the distance between 
signal wires has been chosen $s=(1/10)'=2.54~\text{mm}$.

In a number of articles~\cite{Walenta}-\cite{Erin} the possibility of 
drift chamber with small drift distance ($\sim1-3~\text{mm}$) operation in
high gas gain mode ($\sim10^7$) and high counting rate 
($\sim10^7~\text{sec}^{-1}~\text{cm}^{-2}$) was shown. Drift chamber operation 
in high gas gain mode has a number of advantages. Big amplitude of signal 
enables to exploit readout electronics with high threshold as well as short 
pulses with rapid front improve the time properties of a chamber.

Next in this section the analysis of the cell geometry and gas mixture,
performed by the use of GARFIELD package~\cite{Garfield} are given.

\subsection{Drift cell geometry}
\begin{figure}[ht]
\begin{center}
  \includegraphics[width=8.5cm]{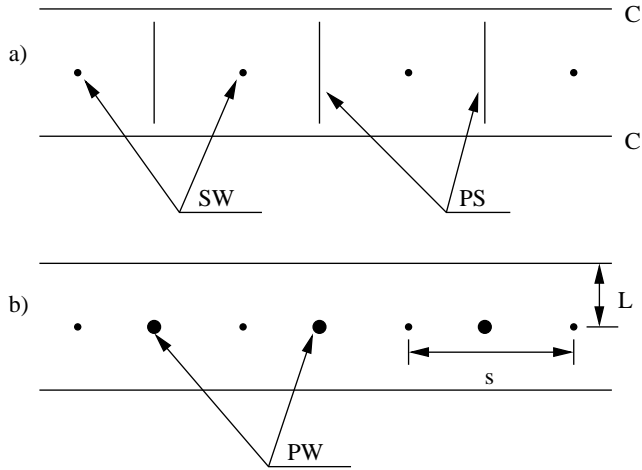} 
  \caption{Schematic layouts of the micro drift chamber electrodes: 
           a) with potential wires PW $d_{pw}=100~\mu\text{m}$ in diameter;
           b) with potential strips $d_{ps}=1.6~\text{mm}$ width;
           SW - anode wires, $d_{sw}=50~\mu\text{m}$ in diameter; 
           C - cathodes, Mylar foil, $20~\mu\text{m}$ thick, 
           $s=2.54~\text{mm}$ - wire spacing, 
           $L=1~\text{mm}$ - gap between sensitive and cathode plane.} 
  \label{fig:cells}
\end{center}
\end{figure}

We considered two possible layouts of MDC cell (fig.~\ref{fig:cells}).
It is supposed, that anode wires are under the zero, cathode planes and
field formative electrodes under the common negative potential.

\begin{figure}[ht]
\begin{center}
  \includegraphics[width=8.5cm]{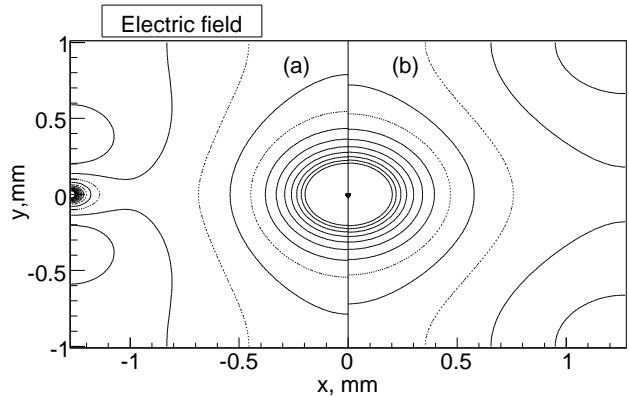} 
  \caption{Electric field map inside different drift cell layout:
           a) cell with potential wires;
           b) cell with potential strips,
           calculated by GARFIELD package.
           The chamber voltage $U=2.5~\text{kV}$. Contours correspond
           to the field intensities 2.5, 5.0, ... 30.0~kV/cm.}
\label{fig:field}
\end{center}
\end{figure}

In fig.~\ref{fig:field} one can see an electric field map inside
different drift cell layouts. The cell with potential wires has rather 
great regions with low value of electric field, which make worse
the time properties of the chamber. Other disadvantage of this layout
is the possibility to hit also an adjacent anode wire in case of particle 
crosses the sensitive plane near potential wire. This is not desirable
in high counting rate condition.

\subsection{Gas mixture}
\begin{figure}[ht]
\begin{center}
  \includegraphics[width=8.5cm]{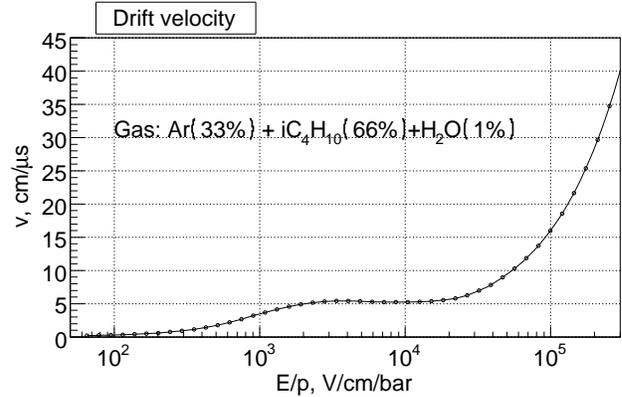} 
  \caption{Drift velocity dependence on electric field, calculated 
           by GARFIELD package.} 
\label{fig:drvel}
\end{center}
\end{figure}

We choose for the Micro Drift Chambers the mixture:  
$\text{Ar}(33\%) + i\text{C}_{4}\text{H}_{10}(66\%) 
+ \text{H}_{2}\text{O}(1\%)$.
This gas mixture is used in DIRAC Drift Chambers system. 
It has the next preferences:
\begin{itemize}
\item
stable work in high gas gain mode;
\item
wide plateau ($>1~\text{kV}$) of efficiency;
\item
fast leading-edge time of pulses ($\sim5~\text{ns}$);
\item
weak drift velocity $v(E/p)$ dependence on electric field $E/p$
practically for whole field value interval inside the drift cell.
(fig.~\ref{fig:drvel}).
\end{itemize}

For this gas mixture at normal pressure the mean number of primary
ionization clusters generated by charged pion with momentum 
$2~\text{GeV}/c$ is $\sim 17$. If particles cross the cell near the 
anode wire then fluctuations of primary ionization spread significantly 
drift times. This spread could be reduced by increasing the pressure 
inside the chamber. As it will be shown below increase of pressure 
in addition leads to more stable chamber operation. 

\begin{figure}[ht]
\begin{center}
  \includegraphics[width=8.5cm]{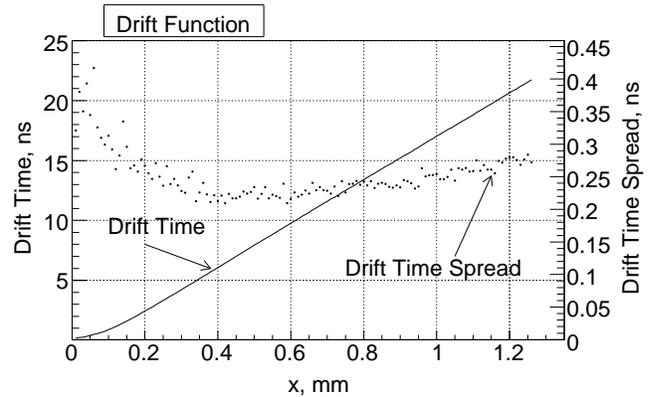} 
  \caption{Drift time and its spread, calculated by 
           GARFIELD package for MDC at $U=3.5~\text{kV}$}
\label{fig:drift}
\end{center}
\end{figure}

In fig.~\ref{fig:drift} the results of drift function simulation for
the selected MDC layout and gas mixture at pressure $p=2~\text{bar(a)}$ 
are shown. As one can see, the drift function is close to linear, 
maximum drift time is $t_{max}=22~\text{ns}$ and the spread of the drift 
times provided by primary ionization fluctuations, diffusion and other
processes in the gas does not exceed $0.5~\text{ns}$.

\section{One cell prototype tests of the Micro Drift Chamber}
To study properties of the Micro Drift Chamber the one cell 
prototype was produced with potential strips (fig.~\ref{fig:cells}b). 
It consists of a signal beryllium bronze wire of diameter 
$d_{aw}=40~\mu\text{m}$, potential strips and cathode planes
made of Mylar foil $25~\mu\text{m}$ thick coated by carbon with 
conductivity $400~\text{Ohm}/\Box$. The drift cell dimensions are 
$2.5\times~2\times~80~\text{mm}^3$. The prototype was placed into the
pressurized box to vary the pressure inside the cell.

We used double scintillation counter telescope and collimated 
radioactive source $^{90}\text{Sr}$ to measure chamber characteristics.

Efficiency and single counting rate curves of MDC cell were measured
at different values of particle flux, pressure and threshold of 
electronics. Also time spectra were measured.

Particle flux was defined by distance between the radioactive source 
and collimator and varied within: 
$(1\div50)\times10^3~\text{sec}^{-1}\text{cm}^{-1}$.

The hit wire signals after preamplifier went to the discriminator and then
to the coincidence circuit, where  scintillation counter signals also came,
and to the counter. Thresholds of discriminator for MDC signal was:
$I_{th}=5, 10, 20$ and $50~\mu\text{A}$.

\begin{figure}[ht]
\begin{center}
  \includegraphics[width=8.5cm]{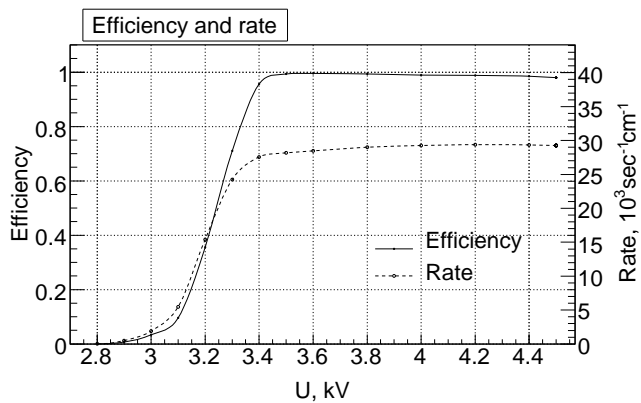} 
  \caption{Efficiency and single counting rate of one MDC cell
           at threshold of electronics $I_{th}=10~\mu\text{A}$.}
\label{fig:eff_cell}
\end{center}
\end{figure}

The chamber was filled with gas mixture:
$\text{Ar}(33\%) + i\text{C}_{4}\text{H}_{10}(66\%) 
+ \text{H}_{2}\text{O}(1\%)$.
The measurements were performed at three values of absolute pressure
inside the chamber: $p=1.0, 1.5, 2.0~\text{bar}$. It was found that
the pressure increasing leads to more stable operation of the detector.
The efficiency plateau expanded, pulse shape became better and drift
times spread was reduced.

In fig.~\ref{fig:eff_cell} typical efficiency and single counting curves
measured at pressure $p=2.0~\text{bar}$ and threshold of discriminator
$I_{th}=10~\mu\text{A}$ are shown.

Also the MDC time spectra were obtained at different values of pressure, 
threshold of electronics and voltage corresponded to efficiency plateau 
at given threshold. Measured spectra have rather complicated shape, which
could be explained by the impossibility to provide the even irradiation
of the drift cell due to low energy of $^{90}\text{Sr}$ electrons. 
Therefore the time spectra is not shown here, because their shape don't
provide information about drift function form. At once the width of the
spectra is around $20~\text{ns}$, which corresponds to the maximum drift 
time calculated by GARFIELD package.

\section{Conclusion}
In this work we have studying  the properties of the drift chamber 
with small drift distance prototype operating in high gas gain mode 
from point of view their application as a coordinate detector in 
the upstream part of the DIRAC setup.

Taking into account the experimental requirements and GARFIELD
simulation results the optimal chamber layout and gas mixture were
chosen. The one cell prototype of the drift chamber was produced
and its characteristics were measures. These measurements were
performed with $^{90}Sr$ radioactive source at various gas pressure 
inside the drift cell, threshold of the discriminator and particle 
flux through the drift cell.

It was shown that the chamber operates more stable at increased pressure.
At $p=2~\text{bar}$, counting rate per wire unit of length up to
$5\times10^{4}~\text{sec}^{-1}\text{cm}^{-1}$ and threshold 
$50~\mu\text{A}$ the efficiency plateau width is few hundreds volts.

The data obtained equally with expected coordinate resolution
make it possible to employ the Micro Drift Chamber for a number 
of applications including the DIRAC experiment.

The next step directed to development of the precise vertex detector
for the DIRAC experiment is the double-plane prototype production
and testing on the beam to check the stability of the chamber in hard 
radiation condition and estimation of the main characteristics of the 
detector: efficiency, coordinate resolution and double track 
resolution.

\end{document}